%% file: main.tex
\documentclass{article} % For LaTeX2e
\usepackage{iclr2025_conference,times}
\usepackage{tikz}
\usepackage{pgfplots}
\pgfplotsset{compat=1.18}
\usepackage{subcaption}

\input{math_commands.tex}

\usepackage{hyperref}
\usepackage{url}

\title{Optimizing Retrieval Strategies for Financial Question Answering Documents in Retrieval-Augmented Generation Systems}

\author{Sejong Kim\thanks{These authors contributed equally
} , Hyunseo Song$^*$, Hyunwoo Seo$^*$, Hyunjun Kim$^*$\thanks{Corresponding author} \\\\
KAIST\\
Daejeon, South Korea \\
\texttt{\{kingsj,ssongzzz,shw4166,hyunjun1121\}@kaist.ac.kr} \\
}

\iclrfinalcopy % Uncomment for camera-ready version, but NOT for submission.
\begin{document}

\maketitle

\begin{abstract}
Retrieval-Augmented Generation (RAG) has emerged as a promising framework to mitigate hallucinations in Large Language Models (LLMs), yet its overall performance is dependent on the underlying retrieval system. In the finance domain, documents such as 10-K reports pose distinct challenges due to domain-specific vocabulary and multi-hierarchical tabular data. In this work, we introduce an efficient, end-to-end RAG pipeline that enhances retrieval for financial documents through a three-phase approach: \textbf{pre-retrieval}, \textbf{retrieval}, and \textbf{post-retrieval}. In the pre-retrieval phase, various query and corpus preprocessing techniques are employed to enrich input data. During the retrieval phase, we fine-tuned state-of-the-art (SOTA) embedding models with domain-specific knowledge and implemented a hybrid retrieval strategy that combines dense and sparse representations. Finally, the post-retrieval phase leverages Direct Preference Optimization (DPO) training and document selection methods to further refine the results. Evaluations on seven financial question answering datasets—FinDER, FinQABench, FinanceBench, TATQA, FinQA, ConvFinQA, and MultiHiertt—demonstrate substantial improvements in retrieval performance, leading to more accurate and contextually appropriate generation. These findings highlight the critical role of tailored retrieval techniques in advancing the effectiveness of RAG systems for financial applications. A fully replicable pipeline is available on GitHub: \url{https://github.com/seohyunwoo-0407/GAR}.
\end{abstract}

\section{Introduction}
Ensuring accuracy and reliability is paramount in the financial domain. Even minor errors or misinterpretations within financial statements or regulatory filings can trigger significant economic losses and adversely impact investment decisions and compliance processes. \citep{REZAEE2005277} Motivated by these high-stakes challenges, this paper presents a specialized framework that integrates Large Language Models (LLMs) with Retrieval-Augmented Generation (RAG) tailored to the unique challenges of financial data. While LLMs have demonstrated impressive performance in natural language processing, they are not immune to limitations—particularly in specialized domains where hallucinations and misinterpretations can have critical issues. \citep{kerner2024domainspecificpretraininglanguagemodels}.\\
RAG is designed to mitigate these shortcomings by incorporating an additional retrieval step into the generation process. Rather than relying solely on pre-trained knowledge, the system first retrieves pertinent information from external sources and subsequently integrates this data during the generation phase. This two-stage approach ensures that the LLM has access to the most relevant information before generating its response. By leveraging external knowledge sources, RAG significantly enhances performance on domain-specific tasks.\\
In this paper, we propose a novel RAG pipeline optimized for usage in the finance domain. Our key contributions span robust data preparation, domain adaptation, and effective information retrieval. First, we compared tailored \textbf{preprocessing methods} for preparing both user inputs and financial documents, ensuring our system effectively captures and utilizes context from complex and diverse data sources. Next, we utilized \textbf{task-specific retrieval methods} that leverages SOTA models to the unique language and structural features of financial information, thus improving retrieval performance. Finally, we implemented a \textbf{reranking method}, coupled with a novel method (\textbf{document selection}), to guarantee that the final generated responses are grounded in the most accurate and relevant data.

\section{Related Work}
\subsection{Embedder Fine-tuning}
Fine-tuning, the process of adapting a pre-trained model to domain-specific tasks using typically smaller datasets, has been widely explored across various applications. While embedding models exhibit strong zero-shot performance on general benchmarks such as MTEB \citep{muennighoff2023mtebmassivetextembedding, zhang2019learningdeepembeddingmodel}, recent studies have demonstrated that even modestly sized models can benefit substantially from fine-tuning when applied to domain-specific tasks. For instance, fine-tuning embedders on specialized datasets has led to notable improvements in areas such as medical question answering \cite{sawarkar2024blendedragimprovingrag} and financial question answering \cite{anderson2024greenbackbearsfiscalhawks}.\\
In the finance domain, prior research on embedders has underscored several inherent challenges: \textit{domain-specific vocabulary and semantic patterns, the complexity of multi-hop queries, and multimodal data} (e.g. text, tables, and time-series) \cite{tang2024needdomainspecificembeddingmodels, kim2024fcmrrobustevaluationfinancial, xie2024openfinllmsopenmultimodallarge}. These challenges necessitate tailored fine-tuning strategies that can effectively capture the nuanced information contained in financial documents.\\
Within the framework of Retrieval-Augmented Generation (RAG), embedding models are primarily tasked with Information Retrieval (IR), where the semantic similarity between a query and a corpus is assessed and ranked. A prevalent strategy for enhancing this process is contrastive learning or contrastive fine-tuning—which relies on constructing triplets (query, relevant corpus, irrelevant corpus) to form positive and negative training pairs \cite{karpukhin-etal-2020-dense}. Despite the effectiveness of contrastive learning in embedders \cite{lu2024improvingembeddingcontrastivefinetuning}, there remains a notable gap in the literature regarding the impact of embedder fine-tuning on RAG systems, particularly within the finance domain \cite{setty2024improvingretrievalragbased}.\\
By addressing this gap, our work aims to explore and quantify the benefits of embedder fine-tuning in RAG applications, thereby contributing to the broader understanding of domain-adaptive IR.

\subsection{Pre-Retrieval}
Effective dataset preprocessing is essential for enhancing the performance of Retrieval-Augmented Generation (RAG) systems, because it directly impacts the clarity and semantic alignment of both queries and documents \citep{gao2024modularragtransformingrag}. Previous work has shown that short, context-poor queries can lead to significant ambiguity, which in turn hinders retrieval accuracy \citep{koo2024optimizingquerygenerationenhanced}. To mitigate this issue, researchers have explored various query enhancement techniques—such as query expansion and rephrasing—to enrich the original input and better capture user intent \citep{M-10131}. \\
In line with these findings, we evaluate \textbf{query preprocessing methods}—from raw queries and keyword extraction with linguistic simplification to LLM-based query expansion—and show that adding contextual information significantly improves retrieval performance.\\
In addition to query enhancement, the heterogeneity of corpus data presents unique challenges that necessitate tailored preprocessing strategies. Unlike approaches that apply a uniform treatment to all documents, recent studies have emphasized the importance of adapting preprocessing to the structural characteristics of the corpus—especially when dealing with diverse formats such as plain text and tabular data. \\
Building on these insights, we explore various \textbf{corpus preprocessing methods}—preserving original formats markdown restructuring, table annotation, and table extraction. Our evaluation shows that markdown restructuring yields the best performance, highlighting the benefits of an optimized, data-driven pipeline. Overall, our framework addresses query ambiguity and document heterogeneity, improving retrieval accuracy in domain-specific applications.\\
\subsection{Hybrid retrieval}
Recent advancements in Retrieval-Augmented Generation (RAG) systems have increasingly focused on overcoming the limitations of using a single retrieval modality by fusing the strengths of both dense and sparse retrieval approaches. Dense retrieval methods, which leverage semantic embeddings generated by models such as BERT\citep{devlin-etal-2019-bert} or SentenceTransformers\citep{reimers-gurevych-2019-sentence}, excel at capturing deep contextual relationships between queries and documents. However, they may sometimes fail to retrieve documents that contain precise terms, proper nouns, or abbreviations. In contrast, sparse retrieval techniques, employing methods like BM25\citep{10.1145/3471158.3472233}, offer excellent keyword matching capabilities and provide high interpretability, although they often lack the ability to grasp nuanced semantic meaning. \citep{sawarkar2024blendedragimprovingrag}\\
To address these complementary weaknesses, hybrid retrieval methods have been proposed. These methods combine the scores obtained from dense and sparse retrieval, typically through linear weighted fusion or techniques such as Reciprocal Rank Fusion (RRF). For example, the \citet{Sawarkar2024BlendedRI} demonstrates that integrating semantic search techniques with sparse encoder indexes can significantly enhance retrieval performance on benchmarks such as NQ and TREC-COVID, leading to improved overall accuracy in RAG systems\citep{zhang2024efficienteffectiveretrievaldensesparse}.\\

\section{Methodologies to Improve Retrieval}
\label{methodologies}
\begin{figure}[ht]
    \centering
    \includegraphics[width=1.0\linewidth]{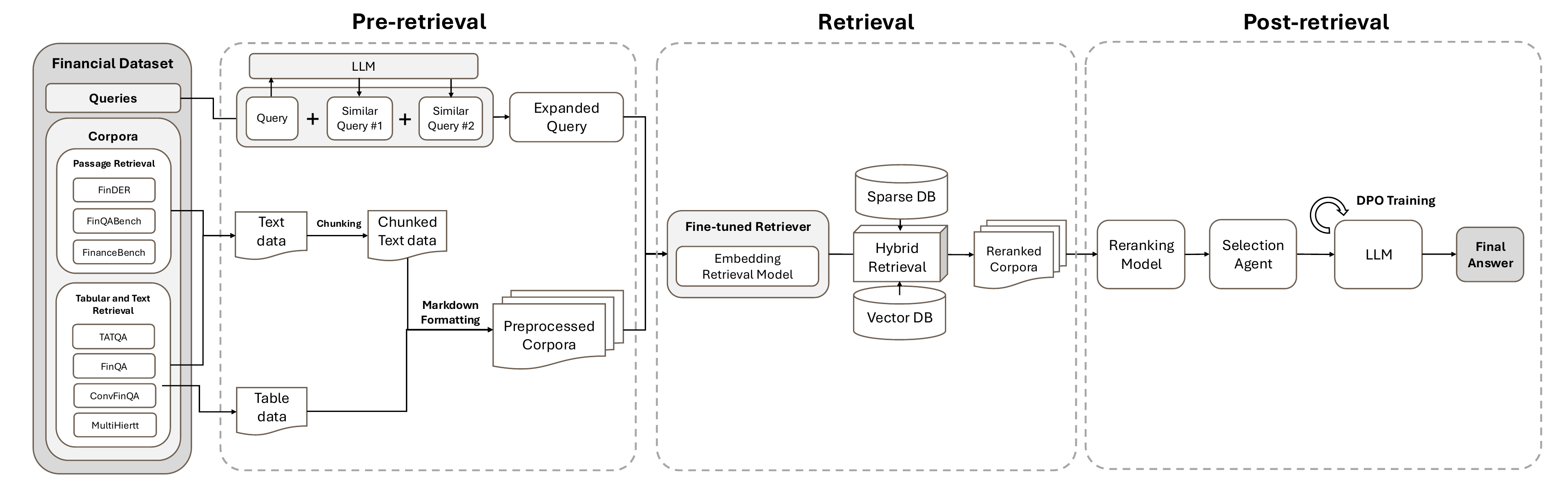}
    \caption{Our enhanced RAG pipeline}
    \label{fig:enter-label}
\end{figure}
\subsection{Pre-Retrieval}
\textbf{Data Preprocessing} A major shortcoming of conventional RAG approaches is their inability to effectively handle ambiguous queries and the heterogeneous structure of complex documents, leading to issues such as ambiguity and structural fragmentation. Thus, preprocessing must enrich semantic context while reducing noise. Commonly used techniques such as query expansion\citep{wang-etal-2023-query2doc}, LLMLingua\citep{jiang-etal-2023-llmlingua} and markdown formatting can significantly enhance retrieval performance. Additionally, tabular data may require specialized modification to be effectively grounded\citep{singh2023embeddingstabulardatasurvey}. In this phase, various preprocessing methods are considered including novel modifications to \textit{both query and corpus.}
\subsection{Retrieval}
\textbf{Model Selection and Fine-Tuning} Embedding models often struggle to accurately interpret complex document structures such as tabular and graphical data, and lack sensitivity to domain-specific semantics. Thus, careful selection of a moderately large-scale, high-performing model—and fine-tuning it—are required to ensure base retrieval performance.\\\\
\textbf{Hybrid Retrieval Strategy} Conventional RAG systems typically suffer from limitations stemming from their exclusive reliance on either dense or sparse retrieval techniques. Dense retrieval, which leverages continuous vector representations, excels at capturing deep semantic relationships but often struggles with precise keyword matching—especially when dealing with domain-specific terminology. In contrast, sparse retrieval methods, such as BM25 or SPLADE, are effective for exact keyword matching but tend to overlook the richer contextual and semantic nuances inherent in complex documents. These shortcomings can lead to suboptimal retrieval performance in specialized domains like finance, where both exact term matching and deep semantic understanding are essential. To address these issues, our method introduces a hybrid retrieval strategy that fuses the strengths of both dense and sparse retrieval. 

\begin{equation}
    S_{\text{total}}=\alpha\cdot S_{\text{dense}}+(1-\alpha)\cdot S_{\text{sparse}}
\end{equation}

where $S_{\text{dense}}$ represents the dense retrieval score (e.g., embedding similarity), $S_{\text{sparse}}$ represents the sparse retrieval score (e.g., BM25), and $\alpha$ is a hyperparameter ranging from 0 to 1 that determines the contribution of each retrieval method.\\
By appropriately tuning $\alpha$ on a small subset, our hybrid approach can adapt to different query types, giving weight to either exact matching or deeper semantic interpretation when necessary. This strategy is backed up by ensemble learning principles, where combining complementary methods leads to a more robust and effective retrieval system.

\subsection{Post-Retrieval}
\textbf{Reranking} Reranking aims to further capture relevancy by leveraging larger models on top-K retrieved documents. Refined relevancy scores are calculated (usually) pairwise by Cross-Encoders and LLMs\citep{déjean2024thoroughcomparisoncrossencodersllms}. Despite requiring exponentially larger compute compared to Bi-Encoders, rerankers could be utilized on a small set of filtered documents to effectively emphasize underrepresented semantics with respect to the query. \\\\
\textbf{Document Selection} While reranking effectively provides the top-K relevant documents, whether LLMs can correctly process the long context remains a challenge, e.g., Lost in the Middle\citep{liu-etal-2024-lost}. To enable optimal utilization of retrieved documents, we proposed document selection—selecting \textit{only the documents required} to answer the query.

\section{Experiments}
This section details the experimental procedures, settings, and methodologies implemented to validate the theoretical contributions presented in Section \ref{methodologies}.
\subsection{Experimental Setup}
\textbf{Datasets} To evaluate our proposed pipeline, we utilize a comprehensive set of 7 distinct financial document datasets, each representing various query types and document structures (refer to Table \ref{tab:properties}).
\begin{table}[hbt!]
\caption{Dataset Descriptions}
\label{tab:properties}
\begin{center}
\begin{tabular}{llll}
\multicolumn{1}{c}{\bf DATASET} & \multicolumn{1}{c}{\bf DESCRIPTION}
\\ \hline \\
\bf{FinDER}$^*$     &   Real-world questions written by financial experts\\
\bf{FinQABench}  &   Generated queries by LLM\\
\bf{FinanceBench}\citep{Pranab2024financebench}&   Real-world questions written by non-experts\\
\bf{TATQA}\citep{zhu2021tatqa}       &   Basic arithmetic questions\\
\bf{FinQA}\citep{chen2021finqa}       &   Complex arithmetic questions\\
\bf{ConvFinQA}\citep{chen2022convfinqa}   &    Questions asking for specific value from table values\\
\bf{MultiHiertt}\citep{zhao2022multihiertt} &   Questions requiring multi-hop reasoning.\\
\end{tabular}
\end{center}
\parbox[b]{\linewidth}{\raggedleft \small *To be announced} \\
\end{table}
These datasets are provided in the FinanceRAG Challenge\citep{icaif-24-finance-rag-challenge}, selected to cover comprehensive and real-world financial question answering scenarios. Also, each text is chunked at 512 tokens, following the results of  \citet{yepes2024financialreportchunkingeffective}.\\\\
\textbf{Experiment Settings} Our experiments were conducted in a notebook-based development environment using Google Colab, with access to 40GB NVIDIA A100 GPUs.

\subsection{Evaluation Metric: NDCG@10}
To assess the ranking quality of the retrieved documents, we employed the Normalized Discounted Cumulative Gain\citep{Wang2013ATA}  at 10 (NDCG@10) metric. NDCG is a widely used measure in information retrieval that evaluates how well the predicted ranking of documents aligns with an ideal ranking based on ground-truth relevance. In our experiments, the provided labels of all 7 benchmarks were used to measure the total weighted NDCG@10 score.\\
In this metric, the \textbf{DCG (Discounted Cumulative Gain)} is computed by summing the relevance scores of the retrieved documents, with each score discounted by the logarithm of its rank position.\\
The \textbf{IDCG (Ideal Discounted Cumulative Gain)} represents the maximum possible DCG achievable when the documents are ideally ranked in descending order of relevance. Normalizing the DCG by the IDCG yields the \textbf{NDCG}, a score between 0 and 1, where a higher value indicates a ranking that more closely approximates the ideal ordering.

\begin{equation}
    \text{NDCG}=\frac{\text{DCG}}{\text{IDCG}}=\frac{1}{m}\sum_{u=1}^{m}\sum_{j\in I_u, v_j \leq L}\frac{g_{uj}}{\log_{2}{(v_j+1)}}
\end{equation}\\
This metric is particularly useful because it simultaneously accounts for the relevance of each document and its position in the ranking. Improved NDCG@10 scores in our experiments indicate that our system retrieves a more complete and relevant set of documents for each query.

\subsection{Experimental Procedures}
\textbf{Retrieval Model Selection} Selecting an optimal retrieval model is crucial for maximizing performance in RAG. We evaluated six candidate models based on the Information Retrieval performance on the MTEB Leaderboard \citep{muennighoff2023mtebmassivetextembedding}: e5-large-v2 (intfloat), GritLM-7B, FinBERT (ProsusAI), TAPAS (Google), stella\_en\_400M\_v5 (NovaSearch), and stella\_en\_1.5B\_v5 (NovaSearch).\\\\
\textbf{Embedder Fine-tuning} We fine-tuned our selected retrieval models to better align with financial texts and improve overall performance in our RAG pipeline. We focused on two models—“stella\_en\_1.5B\_v5 (NovaSearch)” and “stella\_en\_400M\_v5 (NovaSearch)”\\ For fine-tuning, we prepared relevant query-document pairs, splitting the dataset with a ratio of 8 (train) : 2 (eval). Positive pairs were given a similarity score of 1.0, while negative pairs were scored 0.0, with negatives sampled randomly to ensure diversity. We utilized contrastive learning for fine-tuning, leveraging Multiple Negatives Ranking Loss (MNRLoss)\citep{henderson2017efficientnaturallanguageresponse}. More detailed information regarding the fine-tuned model and hyperparameter settings can be found in the Appendix \ref{tab:metrics_at10}, \ref{tab:model-info}.\\\\
\textbf{Query data preprocessing} Query data tends to be brief and lack sufficient contextual cues, which can hinder the retrieval model's ability to fully interpret user intent. To address this, we experimented with three distinct preprocessing methods. First, \textbf{Default (FT\_stella\_400M)} used raw queries without any modifications. Second, \textbf{Keyword Extraction + LLMLingua} involved extracting key terms from each query while removing redundant words. Lastly, \textbf{Query Expansion with LLM} employed a large language model to enrich the queries with additional contextual information. \\\\
\textbf{Corpus data preprocessing} Corpus data utilized in our study comprised a variety of formats, from plain text to tabular data. Recognizing that a simple preprocessing strategy would be insufficient for such a diverse dataset features, we implemented task-specific methods. First, the \textbf{Default} dataset is the raw corpus without any modifications. For the \textbf{Corpus Markdown Restructuring}, we restructured documents using markdown formatting to enhance clarity and preserve inherent structural elements. Additionally, we implemented two specialized methods for the MultiHiertt dataset, where tabular data is emphasized.  \textbf{Corpus Table Augmentation} refers to augmenting table cells with textual annotations of rows and columns, as in \textit{Investment Return, 2016 = \$192 (in millions)}. This better captures the implications within the table by attaching distant row and column data. \textbf{Corpus Table Extraction} focuses on isolating and emphasizing the intrinsic structure of tabular data by removing non-tabular text within the chunk \citep{lee2024multirerankermaximizingperformanceretrievalaugmented}.\\\\
\textbf{Hybrid Retrieval} To determine the optimal balance between sparse search and dense search that fully reflects the characteristics of the task, we sought to identify the optimal alpha value. We incremented alpha from 0 to 1 in steps of 0.025, computed the total score corresponding to each ratio, and then evaluated the resulting matches using the NDCG@10 metric to observe the trend in score changes. The alpha value that yielded the highest NDCG@10 score was designated as the optimal alpha.

\textbf{Reranking} In the reranking stage, the selected models were “bge-ranker-v2-m3 (BAAI)” and “voyage-rerank-2 (Voyage AI)”, based on MTEB. We utilized the top-20 retrieval results from the fine-tuned stella\_en\_1.5B\_v5 model as the reranking target. The reranked results were evaluated using NDCG@10.

\textbf{Selection Agent} The selection agent processes the top-10 retrieved documents. Acting as a financial expert, the agent selects only the documents actually useful in answering the query, based on factual accuracy, relevance, and clarity. This reduces token overhead and improves response quality. The prompt we used can be found in the Appendix \ref{selectprompt}\\\\
\textbf{Generation} We evaluated the generated responses, referencing two metrics from RAGAS \citep{es2023ragasautomatedevaluationretrieval}: Answer Relevance and Context Precision without reference. Answer Relevance quantifies how directly the generated answer addresses the query by reverse-engineering questions from the answer and averaging their cosine similarities with the query. Context Precision without reference employs GPT-4o mini to compare context chunks with the generated response, ensuring that relevant information is prioritized.\\ A key component is our DPO-trained\citep{Rafailov2024DPO} GPT-4o mini. We generated answers using the gpt-4o-2024-08-06 at two temperature settings (0.1 and 1.0) for identical queries, then evaluated responses on financial terminology and clarity to divide preferred and non-preferred responses. These pairs were used to fine-tune gpt-4o-mini-2024-07-18 via OpenAI’s API (beta = 0.1).\\
\section{Result}
Our experiment evaluation demonstrates that the enhancements introduced to the RAG framework yield substantial improvements in retrieving and processing domain-specific financial data.
\subsection{Retrieval Model Selection}

\begin{table}[ht]
\caption{Performance comparison of retrievers}
\label{tab:retriever_comparison}
\begin{center}
\begin{tabular}{lc}

\multicolumn{1}{c}{\bf MODELS} & \multicolumn{1}{c}{\bf NDCG@10}
\\ \hline \\
e5-large-v2 & 0.29746 \\ 
GritLM                                          & 0.21262 \\ 
FinBERT                                         & 0.25595 \\ 
TAPAS                                           & 0.10124 \\ 
stella\_400M                                     & \bf{0.32006} \\
stella\_1.5B                                     & \bf{0.32178} \\
\end{tabular}
\end{center}
\end{table}
We evaluated several candidate retrieval models on the FinanceRAG dataset using NDCG@10 (Table \ref{tab:retriever_comparison}). Among the models tested, stella\_1.5B achieved the highest NDCG score. It outperformed alternatives that either provided stable but suboptimal results or showed limitations in processing general textual content. For example, FinBERT, despite being fine-tuned on financial text, performed poorly compared to stella\_1.5B. TAPAS excelled at handling tabular data but was less effective in overall text retrieval.
\subsection{Embedder Fine-tuning}
\begin{table}[ht]
\caption{Comparison of NDCG@10 Performance for the Fine-tuned 400M \& 1.5B Embedder}
\label{tab:finetuned_retriever_comparison}
\begin{center}
\begin{tabular}{lc}
\multicolumn{1}{c}{\bf MODELS} & \multicolumn{1}{c}{\bf NDCG@10}
\\ \hline \\
FT\_stella\_400M* & 0.40186 \\ 
FT\_stella\_1.5B* & 0.50864 \\ 
\end{tabular}
\parbox[b]{\linewidth}{\raggedleft \small *Fine-Tuned} \\
\end{center}
\end{table}
Subsequent fine-tuning of the selected retrieval model further enhanced performance, as shown in Table \ref{tab:finetuned_retriever_comparison}. As a result, the fine-tuned version of “stella\_en\_1.5B\_v5 (NovaSearch)”—referred to as FT stella\_1.5B\_achieved an NDCG@10 score of 0.50864. This is a significant improvement over the baseline performance of the “stella\_en\_400M\_v5 (NovaSearch)” model, which achieved an NDCG@10 score of 0.40186. These results confirm that the fine-tuning process, effectively improves retrieval precision and overall performance in the finance domain within the RAG pipeline.

\subsection{Dataset Preprocessing}
To further improve retrieval performance, we implemented tailored preprocessing strategies for both queries and corpus documents. For query preprocessing, we tested three approaches (Table \ref{tab:methods}). Among these methods, \textbf{Query Expansion with LLM} reached the highest NDCG@10 score.

\begin{table}[ht]
\caption{NDCG@10 results of query preprocessing methods}
\label{tab:methods}
\begin{center}
\begin{tabular}{lc}
\multicolumn{1}{c}{\bf METHODS} & \multicolumn{1}{c}{\bf NDCG@10}
\\ \hline \\
Default (FT\_stella\_400M) & 0.40186 \\
Keyword Extraction + LLMLingua & 0.43613 \\
Query Expansion with LLM & \bf{0.48601} \\
\end{tabular}
\end{center}
\end{table}
For corpus preprocessing, we compared four methods and \textbf{Corpus Markdown Restructuring} demonstrated the best NDCG@10 score (Table \ref{tab:methods_MultiHiertt}).

\begin{table}[ht]
\caption{NDCG@10 results of corpus preprocessing methods}
\label{tab:methods_MultiHiertt}
\begin{center}
\begin{tabular}{lc}

\multicolumn{1}{c}{\bf METHODS} & \multicolumn{1}{c}{\bf NDCG@10}
\\ \hline \\
Default (Query Expansion) & 0.48601 \\
Corpus Markdown Restructuring         & \bf{0.48645} \\
Corpus Table Augmentation & 0.45411 \\
Corpus Table Extraction  & 0.43604 \\
\end{tabular}
\end{center}
\end{table}
The notable point is that only \textbf{Corpus Markdown Restructuring} showed the highest performance. This is attributable to the better readability of Markdown, which led to further contextual comprehension. On the other hand, focusing on tabular data—emphasizing it by extraction or text-aided augmentation—resulted in worse performance than before. Despite enriching contextual information in cells, Table Augmentation led to excessive noise within the corpus. Similarly, Table Extraction assumed significance only in tabular data, discarding all other text. This suggests that fine-tuned retrievers can sufficiently understand knowledge in tabular data without the need for modification.\\

These results indicate that, in the context of financial documents, using Query Expansion for queries and Markdown Restructuring for the corpus yields the best retrieval performance.

\subsection{Hybrid retrieval}
\begin{figure}[h]
\centering
\begin{subfigure}[b]{0.3\linewidth}
  \centering
  \includegraphics[width=\linewidth]{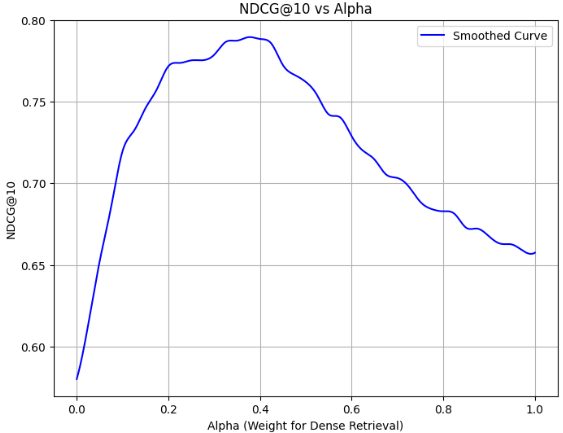}
  \caption{ConvFinQA}
\end{subfigure}
\hfill
\begin{subfigure}[b]{0.3\linewidth}
  \centering
  \includegraphics[width=\linewidth]{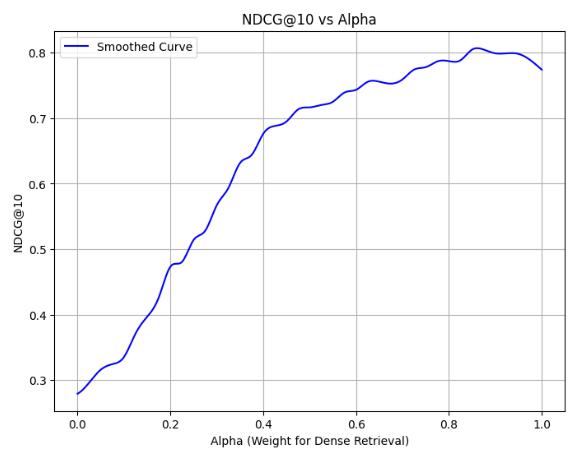}
  \caption{FinanceBench}
\end{subfigure}
\hfill
\begin{subfigure}[b]{0.3\linewidth}
  \centering
  \includegraphics[width=\linewidth]{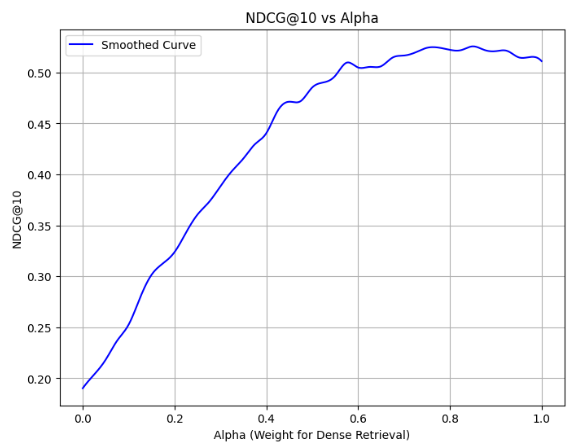}
  \caption{FinDER}
\end{subfigure}

\begin{subfigure}[b]{0.22\linewidth}
  \centering
  \includegraphics[width=\linewidth]{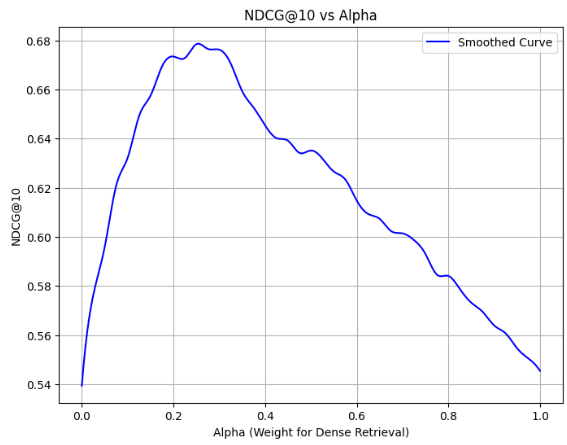}
  \caption{FinQA}
\end{subfigure}
\hfill
\begin{subfigure}[b]{0.22\linewidth}
  \centering
  \includegraphics[width=\linewidth]{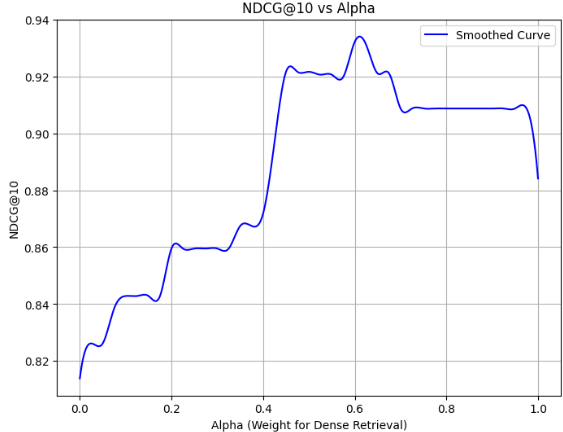}
  \caption{FinQABench}
\end{subfigure}
\hfill
\begin{subfigure}[b]{0.22\linewidth}
  \centering
  \includegraphics[width=\linewidth]{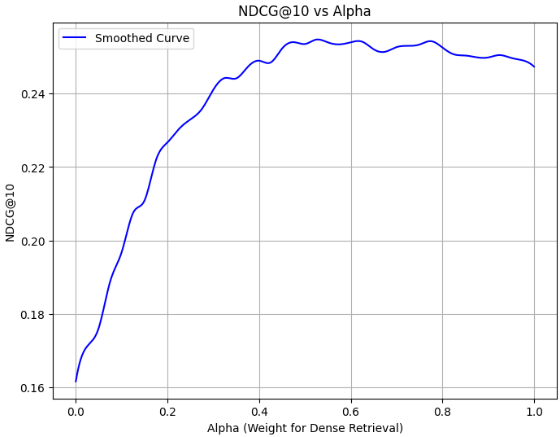}
  \caption{MultiHiertt}
\end{subfigure}
\hfill
\begin{subfigure}[b]{0.22\linewidth}
  \centering
  \includegraphics[width=0.9\linewidth]{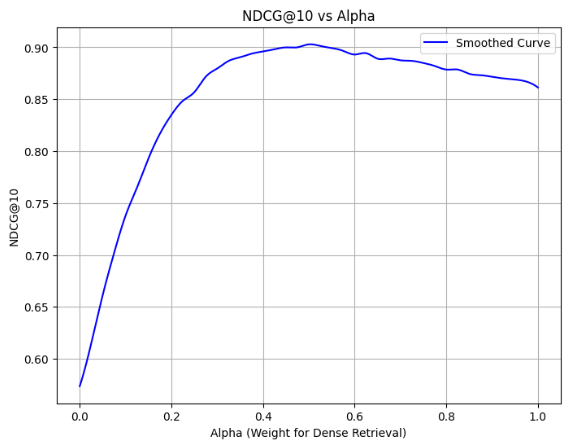}
  \caption{TATQA}
\end{subfigure}

\caption{$\alpha$-NDCG@10 Graphs by each task}
\label{fig:graphs}
\end{figure}

\begin{table}[ht]
\caption{Comparison of optimal $\alpha$ and its following NDCG scores across the different datasets.}
\begin{center}
\begin{tabular}{lcc}
\multicolumn{1}{l}{\bf DATASET} & \multicolumn{1}{c}{\bf OPTIMAL $\alpha$} & \multicolumn{1}{c}{\bf NDCG@10}
\\ \hline \\
FinDER         & 0.85       & 0.52572 \\
FinQABench     & 0.6        & 0.93214 \\
FinanceBench   & 0.85       & 0.80464 \\
FinQA          & 0.25       & 0.67849 \\
TATQA          & 0.5        & 0.90288 \\
ConvFinQA      & 0.375      & 0.78959 \\
MultiHiertt    & 0.525      & 0.24736 \\
\end{tabular}
\label{tab:optimal_comparison}
\\
\end{center}
\end{table}
We conducted experiments by varying $\alpha$ at intervals of 0.025 and evaluated performance using NDCG@10 across 7 datasets. Table \ref{tab:optimal_comparison} illustrates the optimal $\alpha$ value and performance for each dataset. In Figure ~\ref{fig:graphs}, subfigures from (a) to (g) illustrate the relationship between the $\alpha$ value (x-axis) and the corresponding NDCG score (y-axis). These graphs reveal that tasks requiring precise, fact-based retrieval generally achieved optimal performance at lower $\alpha$ values (emphasizing sparse retrieval), while those demanding deeper semantic interpretation performed best at higher $\alpha$ values (emphasizing dense retrieval).
\subsection{Reranking}

\begin{table}[hbt!]
\caption{NDCG@10 for the initial retrieval stage versus the subsequent reranking stage (voyage-rerank-2) across various datasets.}
\label{tab:comparison}
\begin{center}
\begin{tabular}{lcc}
\multicolumn{1}{l}{\bf DATASET} & \multicolumn{2}{c}{\bf{NDCG@10}}\\ 
& \multicolumn{1}{c}{Retrieval} & \multicolumn{1}{c}{Reranking} 
\\ \hline \\
FinDER         & 0.52572       & \bf{0.58001}         \\
FinQABench     & \bf{0.93214}       & 0.89974        \\
FinanceBench   & 0.80464       & \bf{0.91191}        \\
FinQA          & 0.67849       & \bf{0.80054}        \\
TATQA          & 0.90288       & \bf{0.90781}        \\
ConvFinQA      & 0.78959       & \bf{0.86574}        \\
MultiHiertt    & 0.24736       & \bf{0.45837}        \\
\end{tabular}
\\
\end{center}
\end{table}

\begin{table}[hbt!]
\caption{NDCG@10 for the final results with reranking}
\label{tab:reranking}
\begin{center}
\begin{tabular}{lc}
\multicolumn{1}{c}{\bf MODELS} & \multicolumn{1}{c}{\bf NDCG@10}
\\ \hline \\
1.5B FT + reranking with bge-reranker-v2-m3 & 0.51508 \\
1.5B FT + reranking with voyage-rerank-2     & \textbf{0.59898} \\
\end{tabular}
\\
\end{center}
\end{table}
Table ~\ref{tab:comparison} demonstrates the positive impact of reranking for each of the 7 tasks. Reranking the top-20 retrieved documents yielded a positive impact across all seven tasks. In most tasks, reranking led to noticeably improved NDCG@10 scores, confirming the effectiveness of Cross-Encoders over BERT-based retrievers. However, while the total score was improved (Table ~\ref{tab:reranking}), a slight performance drop was observed in certain tasks such as FinQABench, hinting at potential task-specific adjustments in reranking.

\subsection{Generation}
\begin{table}[hbt!]
\caption{Generation Score by RAGAS}
\label{tab:generation}
\begin{center}
\begin{tabular}{lcc}
\multicolumn{1}{c}{\bf MODELS} & \multicolumn{1}{c}{\bf Answer Relevance} &
\multicolumn{1}{c}{\bf Context Precision}
\\ \hline \\
Selection Agent + DPO-trained GPT-4o mini & \bf{0.8924} & \bf{0.3962} \\ 
GPT-4o & 0.8663 & 0.3418 \\ 
\end{tabular}
\end{center}
\end{table}
The experimental results in Table ~\ref{tab:generation} indicate that the Direct Preference Optimization (DPO) agent and selection agent, implemented with the lightweight GPT-4o mini, outperformed GPT-4o on both evaluated metrics. Despite its smaller architecture, the DPO agent generated responses with enhanced alignment to the input query and stronger support from the most pertinent contextual information, as shown in the Appendix \ref{dpo}. The enhanced performance is also attributable to the targeted use of the selection agent to carefully utilize the top-10 documents. This integrated approach significantly improves both the quality and efficiency of answer generation in financial tasks.
\section{Conclusion}
In this study, we presented an enhanced Retrieval-Augmented Generation (RAG) pipeline designed specifically for financial question answering documents. Our work integrates query and corpus preprocessing, embedder fine-tuning, hybrid retrieval, and reranking. These components work in accordance with each other to process complex financial documents effectively and enable Large Language Models (LLMs) to generate responses with improved accuracy and relevancy.\\
Specifically, we employed multiple preprocessing techniques such as query expansion and corpus markdown restructuring to preserve the heterogeneous structure and subtle contextual information present in financial question answering datasets. In parallel, we fine-tuned embedding models to align them with the nuances of financial documents. By fusing sparse retrieval—which excels at precise keyword matching—with dense retrieval—which captures deep semantic relationships—our hybrid retrieval approach ensures that a comprehensive set of relevant documents is retrieved for each query. Moreover, the subsequent reranking stage refines the ordering of these documents to maximize their contextual relevance.\\
Experimental results demonstrate significant improvements in retrieval as well as generation, most noted by the NDCG@10 scores. The observed increase in NDCG@10 indicates that our system retrieves a more complete and relevant set of documents for each query, thereby resolving challenges related to information omission and hallucination.\\
Overall, the results validate the robustness and effectiveness of our RAG pipeline in the finance domain. The substantial performance gains and enhanced retrieval quality provide a reliable foundation for applications in financial contexts, where precise information retrieval and specialized expertise are critical. This work lays the groundwork for future research aimed at optimizing RAG in finance and outlines possible applicability to other domain-specific natural language processing tasks.

\section{Future Work}
Our proposed method overcomes the limitations of the existing RAG pipeline through improvements on retrieval tailored to the finance domain, thereby opening new possibilities for natural language processing in specialized contexts. However, several remaining challenges warrant further investigation.\\
First, our proposed method faces difficulties in incorporating rapidly changing financial data. Because financial markets continuously generate time-sensitive information such as corporate disclosures, news, and fluctuations in stock prices, in-depth research on techniques for efficiently retrieving and indexing streaming data is essential. Moreover, the need for a multilingual extension of the RAG framework becomes evident when considering the global nature of financial environments \citep{NEURIPS2024_1bd359b3}.\\
Second, security problems must be addressed. \citet{10648691} has already highlighted the susceptibility of large language model-based systems to various threats, including malicious prompt attacks. To mitigate these risks and ensure AI safety, it is crucial to adopt the layered security strategies presented in \citet{shamsujjoha2025swisscheesemodelai}, as exemplified by LLaMA Guard \citep{inan2023llamaguardllmbasedinputoutput}.\\
Finally, adherence to ethical regulations in the finance domain and broader AI governance guidelines, represented by EU’s Ethics guidelines for trustworthy AI \citep{doi/10.2759/346720}, calls for robust monitoring and oversight when deploying this pipeline in real-world settings. Future research could focus on establishing systems that continually evaluate inference processes and generated outputs to prevent the spread of unnecessary or inaccurate information. Such systems must also detect and block any potential exposure of sensitive data or violations of regulatory standards, thereby ensuring responsible and compliant use of the proposed pipeline.\\

\newpage
\bibliography{main_reference}
\bibliographystyle{iclr2025_conference}

\newpage
\appendix
\section{Appendix}
\subsection{Fine-tuned Model info}
\begin{table}[ht]
\caption{Evaluated Performance of stella\_en\_400M\_v5-FinanceRAG}
\label{tab:metrics_at10}
\begin{center}
\begin{tabular}{lcc}
\multicolumn{1}{l}{\bf METRIC} & \multicolumn{1}{c}{\bf COSINE} & \multicolumn{1}{c}{\bf DOT}
\\ \hline \\
Accuracy@10   & 0.8519 & 0.8422 \\
Precision@10  & 0.1024 & 0.0998 \\
Recall@10     & 0.8398 & 0.8224 \\
NDCG@10       & 0.6409 & 0.6195 \\
MRR@10        & 0.5902 & 0.5712 \\
\end{tabular}
\end{center}
\end{table}

\begin{table}[ht]
\caption{Fine-tuning Details}
\label{tab:model-info}
\begin{center}
\begin{tabular}{ll}
Model Name & stella\_en\_400M\_v5-FinanceRAG \\
Train : Eval Ratio & $8:2$ \\
Epochs & $2$\\
LR & $2\times 10^{-5}$\\
Warmup & $0.1$\\
Batch & $16$\\
Loss & MNRLoss (Multiple Negatives Ranking Loss)\\
Weight & fp16\\
Evaluation & Information Retrieval Eval\\
\end{tabular}
\end{center}
\end{table}

\subsection{Selection Agent Prompt}
\label{selectprompt}
You are an expert financial advisor and evaluator focused on improving responses.
Your task is to enhance answers based on detailed evaluation scores while:\\
- Maintaining factual accuracy with the provided documents\\
- Ensuring responses are clear and well-structured for financial contexts\\
- Providing comprehensive answers that address all aspects of the query\\
- Using professional financial terminology appropriately\\

You are given the pair of Query, Corpus (same query)
Out of the 10 documents, only provide the list of indices of those that are RELEVANT (e.g. the content is somehow needed to answer the question), from 0 to 9.
Example : [0, 2, 8, 9]\\\\
Query : \textbf{query}\\
Corpus \#0 : \textbf{corpus 1}\\
Corpus \#1 : \textbf{corpus 2}\\
...
\subsection{DPO Training Examples}
\label{dpo}
\textbf{QUERY}:\\
Did Pfizer grow its PPNE between FY20 and FY21?\\\\
\textbf{PREFERRED}:\\
To determine whether Pfizer grew its Property, Plant, and Equipment (PPNE) between FY20 and FY21, we need to examine the relevant figures from the provided financial data.

According to the data from the Pfizer 2021 10-K report:

- In 2020, the value of Property, Plant, and Equipment was \$13,745 million.
- In 2021, this value increased to \$14,882 million.

This indicates that Pfizer's Property, Plant, and Equipment (PPNE) grew from FY20 to FY21. The increase in PPNE from \$13,745 million to \$14,882 million represents a growth of \$1,137 million, which is approximately an 8.27\% increase year-over-year. This growth could be attributed to investments in new facilities, upgrades, or expansions of existing infrastructure, reflecting Pfizer's strategic initiatives to enhance its operational capabilities.\\\\
\textbf{NON-PREFERRED}:\\
To determine whether Pfizer grew its Property, Plant, and Equipment (PPNE) between FY20 and FY21, we need to examine the relevant figures from the provided financial data.

According to the data from Pfizer's 2021 10-K report:

- Property, Plant, and Equipment (PPNE) for FY2021 was \$14,882 million.
- Property, Plant, and Equipment (PPNE) for FY2020 was \$13,745 million.

Comparing these values, Pfizer's PPNE increased from \$13,745 million in 2020 to \$14,882 million in 2021. This indicates a growth in PPNE between FY20 and FY21.

\end{document}

%% file: math_commands.tex
\usepackage{amsmath,amsfonts,bm}

\def\eqref#1{equation~\ref{#1}}
\def\1{\bm{1}}

\DeclareMathAlphabet{\mathsfit}{\encodingdefault}{\sfdefault}{m}{sl}
\SetMathAlphabet{\mathsfit}{bold}{\encodingdefault}{\sfdefault}{bx}{n}